\documentclass[aps,prl,twocolumn,superscriptaddress,groupedaddress]{revtex4}
\usepackage{subfig}
\usepackage{graphicx}  
\usepackage{dcolumn}   
\usepackage{bm}        
\usepackage{amssymb}   
\usepackage{slashed}
\usepackage{graphicx}				
\usepackage{amsmath}
\usepackage{mathtools}
\usepackage{tikz,pgf}
\usepackage{comment}
\usepackage{asymptote}
\usetikzlibrary{arrows,backgrounds}
\usetikzlibrary{fit,scopes,calc,matrix,positioning,decorations.pathmorphing}
\usepackage[all]{xy}
\usepackage{yfonts}

\newcommand{\Bracket}[1]{\ensuremath{\left\langle#1\right\rangle}}

\begin{document}
\title{$H_{0}$ tensions in cosmology and axion pseudocycles in the stringy universe}
\author{Andrei T. Patrascu}
\address{ELI-NP, Horia Hulubei National Institute for R\&D in Physics and Nuclear Engineering, 30 Reactorului St, Bucharest-Magurele, 077125, Romania}
\begin{abstract}
The tension between early and late $H_{0}$ is revised in the context of axion dark matter arising naturally from string theoretical integrations of antisymmetric tensor fields over non-trivial cycles. Certain early universe cycles may appear non-trivial from the perspective of a homology analysis focused on the early universe, while they may become trivial, when analysed from the perspective of a homology theory reaching out to lower energies and later times. Such phenomena can introduce variations in the axion potential that would explain the observed $H_{0}$ tension. 
\end{abstract}
\maketitle
\section{Introduction}
The accumulation of observational data in favour of a physical tension between low and high redshift determinations of the Hubble constant indicates potential new physics beyond the standard model [1], [2], [3]. There are two methods of determining Hubble constant nowadays, one relying on the cosmic microwave background, under the assumption of a cold dark-matter universe with a cosmological constant $\Lambda$ ($\Lambda$CDM) and the other relying on direct measurements from supernovae. These methods were recently highly refined by employing data both from the Planck satellite (for the cosmic microwave background) and from GAIA, a space telescope system providing accurate data for the later method. The Hubble constant inferred from CMB has been calculated to be $H_{0}=67.27\pm 0.60$ $km s^{-1} Mpc^{-1}$ while the supernovae observations insist on a value of $H_{0}=73.52\pm 1.62$ $km s^{-1} Mpc^{-1}$ with the distinction between the distributions related to the two types of observations being significant. Basically this leads us to a tension between the "local" and the "global" measurements of the Hubble constant that can only hardly be considered to result from systematic errors or other data-related biases. It is worth mentioning that while such a discrepancy has been observed in previous data, the interest of the cosmology community peaked when the latest, far more accurate observational results not only re-confirmed the above mentioned tension, but they also made it sharper. While this observational discrepancy has not been in the focus of theoretical model-builders, there have been at least a few attempts to give explanations based on new physics or modified fundamentals. The reconciliation of the theories with observations however is delicate. Neutrinos [4], axions [5], or other moduli-particles [6] have been considered as possible explanations, together with modified dark energy contributions at the early stages of cosmological evolution. In all cases it was not possible to fully acknowledge the evolution of the Hubble constant as observed presently. 
In this article I will provide a new theoretical approach, although based on axion contributions to the cosmological evolution. While axion solutions have been considered before, no analysis relying on the fundamental origin of axions has been made up to now. 
Axions indeed are dark matter candidates that appear as an extension to the standard model of elementary particles and are motivated by the strong CP problem. The strong CP problem refers to the unexpected absence of a CP violating term in Quantum Chromodynamics (QCD) albeit not forbidden by any usual restrictions. Given the experimental observations, if such a CP breaking term were to exist, it would appear to be unjustifiably small leading to a hierarchy problem. The solution to this problem is given by the so called Peccei-Quinn mechanism that involves axial degrees of freedom and an additional (Peccei-Quinn) symmetry broken spontaneously, and hence giving rise to the Axion, and further broken explicitly by various instanton-like mechanisms, providing mass to the newly required Axion. 
As we may notice, while the Axion, and the associated axial degrees of freedom do solve the strong CP problem, their existence in the standard model, together with the symmetry they rely upon is in many ways "ad-hoc". This changes dramatically however if we consider string theory and compactified extra-dimensions. Indeed, once extended objects (like strings or branes) are being considered, and once the background manifold is seen as a higher-dimensional manifold with compact extra-dimensions, we may consider various non-trivial topological cycles emerging, each providing us with axion-like degrees of freedom. From this perspective, not only is the axion a common component that should emerge in any description of high energy physics, but it is also, in a sense, unavoidable. Indeed, many non-equivalent cycles should give rise to their own specific axions, leading to what is known as axion proliferation, or the axiverse. While string theory strongly favours axions, it remains to be seen whether all axions or axion-like particles will have a similar impact on cosmology. As axions appear from the integration of asymmetric tensor fields over non-trivial cycles in string theory, one should pay more attention to these elements of string theory. It has been shown that early axions (and in general, early dark matter contributions) would not be able to explain the tension observed for the Hubble constant, however, it is important to note that the disappearance of certain axion cycles before the recombination period would be able to significantly alleviate the cosmological Hubble constant tension. 
Indeed, this article assumes that the very early universe already was described by a compactified manifold but on this manifold, not all significant cycles were true cycles, some of them being described in terms of pseudo-homology groups, giving them a significant role only at an extremely early stage of the universe. They would dissipate afterwards, becoming basically undetectable in terms of homology groups at the recombination stage, leaving the true cycles associated to axions which we may hope to detect today. Indeed this interplay between homology and pseudo-homology at an early stage of the cosmological expansion may offer an explanation for the observed Hubble tension without implying exotic/esoteric dark energy contributions at the very early cosmological evolution. 
There are several tools we can use to describe this observation, pseudo-homology being one of them, together with the detectability of topological features with homology theories with non-trivial coefficients. While these tools are mathematically accurate, they may not be familiar to the reader, hence I will also give an interpretation using a more standard language originating in string theory. I will also give an example using mostly the language of cosmology. 
\section{WIMPs and Axions in cosmology and particle physics}
Axion like particles are extremely light and weakly coupled degrees of freedom which can be considered dark matter candidates (within certain domains [7]) and are well motivated both within QCD and string theory. On one side the QCD axion is an essential component in the solution of the strong CP problem, on the other side, axion like particles in general emerge naturally from integration over non-trivial cycles in string theory. The sheer quantity of such cycles leads to the so called "axiverse" which contains an axion like particle for every energy decade. 
From a cosmological perspective a dark matter candidate must resolve the missing matter issue observed already in the early observations by Zwicky, and confirmed to almost perfect accuracy in the first and second half of the 20th century. While the postulation of a dark matter particle is obviously required due to cosmological arguments, there are several particularities required by high energy physics models that add a stronger foundation to the dark matter claims. 
On general cosmological grounds dark matter should have the following properties: first, they should be non-relativistic candidates usually represented by massive particles which are not expected to be faster than the average galactic escape velocity, and hence should play a role in explaining the galactic rotation curves observed during the past decades on a vast number of galaxies. 
Second, they ought to be non-baryonic candidates, namely candidates carrying neither electric nor colour charges, and finally, they should be stable enough to ensure their lifetime would reach out to the current age of the universe and also to allow us to expect them to continue with a lifetime many orders of magnitude greater than the life of the universe. 
Dark matter candidates are produced in the early universe either through processes taking place at thermal equilibrium (thermal production) or in processes taking place away from thermal equilibrium (the so called non-thermal production). The thermal production usually appears at the freeze-out temperature and will emerge from relics at thermal equilibrium at this stage of cosmic evolution, or will appear in scatterings and decays of other particles in the original plasma. The non-thermal production involves usually the coherent motion of bosons associated to a bosonic field or from out-of-equilibrium decays of heavier states. Clearly the standard model of elementary particles cannot accommodate such dark matter candidates, the only valid alternative coming from various extensions of the standard model both in the direction of heavier and lighter particles. 
Various observational results excluded most of the common standard model candidates, including the massive, compact, and weakly radiating candidates like black holes, neutron stars, or a potential high density of planetary bodies [8], as well as massive neutrinos excluded by calculations involving their relic abundance. The current dark matter searches focus on the weakly interacting massive particles, also known as WIMPs which represent a broad category of particles usually required by supersymmetry. The gauge hierarchy problem has a simple supersymmetric solution involving the neutralino. The strong-CP problem has another simple solution involving a new Peccei-Quinn symmetry spontaneously (and explicitly) broken, leading to the axion. The axion is a very well motivated non-thermal relic appearing in SUSY models as a supermultiplet containing the axion (a), the spin $\frac{1}{2}$ $R$-parity odd axino ($\tilde{a}$), and the $R$-parity even spin-$0$ saxino ($s$). Their interaction strength is particularly weak, the axion, being the fermionic super-partner of the axion is seen as a WIMP, being on the massive side of the spectrum, but with an extremely weak interaction strength. Its mass is strongly model dependent, and they can be either thermal or non-thermal relics. The axion on the other side, as an example of a non-thermal relic has a interaction strength strongly suppressed by the Peccei-Quinn breaking scale $f_{a}\sim 10^{11}$ GeV.  The interaction strength is, as usually, given by $(\frac{m_{W}}{f_{a}})^{2}$ where $m_{W}$ represents the weak scale. As an additional example, the gravitino, $\tilde{G}$, the SUSY partner of the graviton, is a neutral Majorana fermion with a coupling to ordinary particles strongly suppressed by the Planck scale via $(\frac{m_{W}}{M_{Planck}})^{2}$. Particle relics from the early epochs of the universe can span a enormous range both in mass and in cross section, as they may be generated by very different production mechanisms in the early universe. The WIMP thermal relicts present us with an interesting connection between the cold-dark-matter relic density and the electroweak interaction strength. Because during the early universe stage the WIMPs are considered to be in thermal equilibrium at temperature $T\geq m_{X}$, their number density as a function of time is determined by the Boltzmann equation
\begin{equation}
\frac{dn_{X}}{dt}=-3 H n_{X}-\Bracket{\sigma_{ann}v}(n_{X}^{2}-n_{eq}^{2})
\end{equation}
here $H$ is the Hubble constant which for the radiation dominated universe is given by $H^{2}=\frac{\rho_{rad}}{3M_{Planck}}$, the equilibrium density is $n_{eq}$ and the term $\Bracket{\sigma_{ann}v}$ represents the thermally averaged cross section for the WIMP annihilation times the relative velocity. In the early universe, the number density for WIMPs follows the equilibrium density. As time passes, the temperature reaches a value $T_{fr}$ known as freeze-out point where the expansion rate becomes larger than the annihilation rate and the Hubble term becomes of major importance. After that point, the WIMP's number density in a co-moving volume becomes effectively constant. The present day WIMP relic density can be found as a solution of the Boltzmann equation given by 
\begin{equation}
\Omega_{X}h^{2}\cong \frac{s_{0}}{\rho_{c}/h^{2}}(\frac{45}{\pi^{2} g_{*}})^{1/2}\frac{1}{x_{f} M_{Planck}}\frac{1}{\Bracket{\sigma_{ann}v}}
\end{equation}
where $g_{*}$ is the number of relativistic degrees of freedom at freeze-out, $s_{0}$ is the present day entropy density, and $x_{f}=T_{fr}/m_{X}$ the freeze-out temperature scaled to the WIMP mass. Following ref. [10] and introducing the data from ref. [11] for $s_{0}$, $\rho_{c}$ and $M_{Planck}$ and using the measured value for $\Omega_{X}h^{2}\cong 0.12$ we find 
\begin{equation}
\frac{\Omega_{X}h^{2}}{0.12}\cong \frac{1}{\Bracket{\frac{\sigma_{ann}}{10^{-36}cm^{2}} \frac{v/c}{0.1}}}
\end{equation}
The result of this calculation is interpreted in the sense that a cross section of 1pb and typical WIMP speeds at freeze-out temperature provide the exact present day relic density of dark matter. This is why the WIMP dark matter may be related to new physics which was expected to appear at or around electroweak level. Another motivation for this was the stabilisation of the Higgs boson mass, which will not be discussed here. Needless to say, no new dark matter particles around this scale have so far been detected. We can understand this by thinking that $\sigma_{ann}\sim \frac{g^{4}}{m_{X}^{2}}$ where only the fraction needs be fixed, both $g$ and $m_{X}$ being allowed to vary on relatively broad ranges while still being consistent with the freeze-out mechanism. 
Our concern in this article however will not be with the WIMP dark matter candidates, but instead with the axions which from a cosmological perspective can be regarded as a source for bosonic coherent motion (BCM). The BCM involving the axion implies a light boson with a very long lifetime. As there exists one axion that solves the strong-CP problem, known as the QCD axion, if this is supposed to make up for the dark matter, its mass should be smaller than $24$ eV to be able to exist until the current age. The other axions (also known as axion-like particles, short ALP) are very similar with the QCD axion, arising in a similar way from string theory, with the main distinction that their mass is not linked to the Peccei-Quinn scale $f_{a}$. Such axions are still coupled to the electromagnetic field by means of a term 
$(a_{ALP}/f_{a})F_{\mu\nu}\tilde{F}^{\mu\nu}$. When not bound by the restrictions of the Peccei-Quinn solution of the strong-CP problem, the axion can couple to the QCD anomaly by a term like
\begin{equation}
L=\frac{\alpha_{s}}{8\pi f_{a}}aG^{a}_{\mu\nu}\tilde{G}^{a\mu\nu}
\end{equation}
where the dual gluon field strength is $\tilde{G}^{a\mu\nu}=\frac{1}{2}\epsilon^{\mu\nu\rho\sigma}G^{a}_{\rho\sigma}$ and $\alpha_{s}=g_{s}^{2}/4\pi$ is the strong coupling constant. Such coupling can be obtained by integrating the coloured heavy fields below the Peccei-Quinn breaking scale $f_{a}$ but above the electroweak scale $v_{EW}$. After integrating out all the heavy PQ-charged fields, the axion coupling Lagrangian at low energy in terms of the effective couplings $c_{i}, i=1,2,3$ with the standard model fields is 
\begin{equation}
\begin{array}{c}
L_{int}^{eff}=c_{1}\frac{\partial_{\mu}a}{f_{a}}\sum_{q}\bar{q}\gamma^{\mu}\gamma_{5}q-\\
\\
-\sum_{q}(\bar{q}_{L}mq_{R}e^{ic_{2}a/f_{a}}+h.c.)+\frac{c_{3}}{32\pi^{2}f_{a}}aG\tilde{G}+\\
\\
+ \frac{C_{aWW}}{32\pi^{2}f_{a}}aW\tilde{W}+\frac{C_{aYY}}{32\pi^{2}f_{a}}aY\tilde{Y}+L_{leptons}\\
\\
\end{array}
\end{equation}
The first term involving the derivative interaction proportional to $C_{1}$ preserves the $U(1)$ Peccei-Quinn symmetry. The second term proportional to $c_{2}$ is related to the phase of the quark mass matrix, and the third term, proportional to $c_{3}$ is the anomalous coupling. The coupling between the axions and the leptons is encoded in the interaction term $L_{leptons}$. The axions that are not supposed to represent solutions to the strong-CP problem, namely those which are expected to be particularly light, are described by two types of field theoretical models, one is known as the Kim-Shifman-Vainstein-Zakharov (KSVZ) model, and the other is known as the Dine-Fischler-Srednicki-Zhitnitskii (DFSZ) model. In the first model, at the level of field theory, the axion is present if quarks carry a net PQ charge $\Gamma$ of the global $U(1)_{PQ}$ symmetry. In general, at the standard model level, the six quarks are strongly interacting fermions. The electroweak scale $v_{EW}\cong 246 GeV$ we start taking into account additional, beyond standard model heavy, vectorial quarks $(Q_{i},\bar{Q}_{i})$ but these end up being integrated out from the effective Lagrangian written above. In this model, the only heavy quarks that may appear beyond $v_{EW}$ is must carry PQ charge and hence, below $v_{EW}$ or below the QCD scale $\Lambda_{QCD}$ we have $c_1=c_2=0$ and $c_{3}=1$. The gluon anomaly term given to be proportional with $c_{3}$ is induced by an effective heavy quark loop and solves the strong-CP problem. As a byproduct, the axion field appears as a component of the standard model singlet scalar field $S$. The string axions emerging from $B_{MN}$ are of this type and are defined by the QCD-anomaly coupling at lower energies. These are like the KSVZ axions. 
In the second model one does not introduce any PQ charge in the heavy quark sector beyond the standard model. Instead the standard model quarks are assigned a PQ charge with $c_{1}=c_{3}=0$ and $c_{2}\neq 0$ below the electroweak scale $v_{EW}$. In the same way, the axion is a part of the standard model singlet scalar field $S$. Usually string theory gives also rise to components similar to DFSZ axions in addition to the KSVZ axions. 
The axion has shift symmetry, which is basically just a phase rotation, and the physical observables are invariant under this transformation. Below $f_{a}$ the PQ rotation symmetry is broken into a discrete subgroup which represents the rotation by $2\pi$. This breaking can be seen through the appearance of the $c_{2}$ and $c_{3}$ terms in the Lagrangian. The $c_{2}$ term enters as a phase and a shift by $2\pi$ brings it to the same value, while the $c_{3}$ term is the QCD vacuum angle term, which again, if the vacuum angle is shifted by $2\pi$ comes to the original value. The subgroup corresponding to the common intersection of the subgroups corresponding to $c_{2}$ and $c_{3}$ is preserved. The combination $c_2+c_3$ is invariant under axion shift symmetry and $c_2+c_3$ represents the unbroken discrete subgroup of $U(1)_{PQ}$. This is the domain wall number $N_{DW}=|c_2+c_{3}|$. 

\section{String theory axions}
As noted before, axions appear due to integration of tensor fields over non-trivial cycles arising on the compactified manfiold of string theory. In QCD, the CP-violating term while being a total derivative and hence being trivial from the point of view of classical field equations, has a significant quantum impact due to its non-trivial topological properties. The topologically non-trivial field configurations can be seen by looking at the term in the action 
\begin{equation}
S_{\theta}=\frac{\theta}{32\pi^{2}}\int d^{4}x\epsilon^{\mu\nu\lambda\rho} Tr G_{\mu\nu}G_{\lambda\rho}
\end{equation}
When we shift the parameter $\theta\rightarrow \theta+2\pi$ the action changes by $2\pi$ and hence leaves the partition function unchanged. This suggests that the parameter $\theta$ represents a periodic parameter with a period equal to $2\pi$. The introduction of fermions will bring with it the chiral anomaly and the parameter $\theta$ will have to include the overall phase of the quark mass matrix, modifying it as in 
\begin{equation}
\bar{\theta}=\theta+arg(det(m_{q}))
\end{equation}
However, measurements have shown that $\theta \leq 10^{-10}$. The solution to the strong-CP problem implies making the $\theta$ parameter a dynamical field $a$. At a classical level the action is obviously invariant to any shifts $a\rightarrow a+C$. This means that at the classical level, the axion is the Goldstone boson of a spontaneously broken global symmetry. Quantum perturbative effects preserve this symmetry but non-perturbative, topologically non-trivial QCD field configurations break it explicitly generating a periodic potential for the axion. In the case of the QCD axion, the axion obtains a vacuum expectation value which adjusts itself to render the resulting $\bar{\theta}$ small. The axion couples to the gluons, as noted above, but also to other gauge bosons including photons, and to fermions by means of derivative couplings. It is important to note that their coupling to photons make the axion detectable at extremely intense laser facilities like the ELI-NP [12].

While the justification of the axion resulting from the solution of the strong-CP problem is clear, one may ask more fundamental questions, namely why should a symmetry like the Peccei-Quinn even exist and be explicitly broken by topologically non-trivial QCD fields. Such angular degrees of freedom are certainly unexpected in a fundamental theory based on standard quantum field theory. Pseudoscalars with axion-like properties are however quite natural in string theory compactifications. They may appear as Kaluza-Klein zero modes of antisymmetric tensor fields. The Neveu-Schwarz 2-form $B_{MN}$ that arises in all string theories, or the Ramond-Ramond forms $C_{0,2,4}$ arising in type $IIB$ string theory as well as the $C_{1,3}$ forms arising in $IIA$ string theory are such examples. Higher order antisymmetric tensor fields, upon compactification, typically give rise to a large number of Kaluza-Klein zero modes which are determined by the topology of the underlying compact manifold. In particular, considering a single two form $B_{MN}$ or $C_{MN}$ one obtains a number of massless scalar fields equal to the number of homologically non-trivial closed two-cycles in the underlying manifold. We can look at the Kaluza-Klein expansion for the $B_{MN}$ two-form considering the non-compact coordinates $x$ and the compact coordinates $y$
\begin{equation}
B=\frac{1}{2}\sum b^{i}(x)\omega_{i}(y)+... 
\end{equation}
with $\omega_{i}$ being the basis for closed non-exact two forms (cohomologies) dual to the cycles in our manifold, obeying the constraint that 
\begin{equation}
\int_{C_{i}}\omega_{j}=\delta_{ij}
\end{equation}
Similarly the number of pseudo-scalar zero modes corresponding to $C_{4}$ is equal to the number of homologically non-trivial distinct four-cycles. As it has been noted in [13] the number of cycles in most compactifications is extremely large leading to many axion-like fields being predicted in general by string theory. When going to the four dimensional effective theory the scalar fields resulting from the KK reduction are massless and have a flat zero potential resulting from the higher dimensional gauge invariance of the antisymmetric tensor field action. This invariance also ensures that no perturbative quantum effect can generate a potential. However, antisymmetric tensor fields couple by means of Chern-Simons terms. After KK reduction these terms can couple axion fields to the gauge fields. This has been theoretically observed in type $IIB$ theory with a $C_{2}$ axion with a $D5$ brane wrapped over the associated two-cycle. String theory therefore can produce particles with the qualitative features of the QCD axion. Not only that, but there are quite many such particles expected from string theoretical arguments. 
However, several string axions can be removed at tree level from the string spectrum of light fields by fluxes, branes, or orientifold planes pushing the mass of the axions towards the string scale [13]. Also, even if the axion does not become heavy due to tree level effects, its potential acquires non-perturbative contributions from world-sheet instantons, euclidean D-branes wrapping the cycle, gravitational instantons, etc. Such corrections may ruin the strong CP solution. 

\section{Axion pseudo-cycles and the $H_{0}$ tension}
As observational evidence for a tension between early and late $H_{0}$ accumulates, an explanation based on fundamental physics seems still somehow remote. While the $\Lambda CDM$ model is successful in describing the large scale structure of the universe and is well grounded in the precision observations of the cosmic microwave background by Planck, it seems like local observations of supernovae introduce a tension with the Hubble rate measured from the early cosmic observations, with a statistical significance in several cases already larger than $4\sigma$. 
In this article I will provide a model that will increase the number of parameters to be optimised for cosmological data by introducing a new theoretical model based on pseudo-cycles in string theory. 
The basic idea is as follows: fluctuations in the underlying manifold may generate very early deformations which may provide pseudo-cycles resulting in pseudo-axion fields at a later stage of cosmic evolution. This later stage is still considered early from the perspective of cosmological observations. What basically happens is that such deformations, seen exclusively at high energies will appear as non-trivial cycles corresponding to axion fields which will play a non-trivial role in the very early universe and will vanish in the successive stages, which will still correspond to the early $\Lambda CDM$ stages of the cosmological evolution. The pseudo-cycles correspond to deformed manifolds which are trivial from the perspective of absolute homology theory and hence do not count in the final number of axion fields in an effective theory. However, at intermediate energies, prior to the hot-axion cosmology stage (as presented in [9]) they do have a non-trivial impact as they behave like true axion fields originating from pseudo-cycles perceived as true cycles in the very early stages of cosmic evolution. The length of the deformation towards lower energy (later time) stages represents a new parameter that contributes in the desired way towards the alleviation of the observed $H_{0}$ tension.
The real axion cycles will become relevant later on, inducing the effects well known from the hot-axion model, but this time strongly modulated by the earlier disappearance of the "fake" axions resulting from early pseudo-cycles.

 This gives a mechanism for the "exotic" early dark energy presented in [14] however, it is well grounded in string phenomenology and does not rely on a "mysterious" early dark energy, but instead on a mechanism involving (pseudo) axions which are expected to be detected by subsequent radiative emissions. 
 As presented before, axion fields can be regarded as results of non-trivial cycles arising in the process of compactification. However, the extreme environment in the early universe offers sufficient opportunities for non-trivial topological effects to take place. Among these, we also have a possible local deformation of the underlying manifold that would result in more or less extended pseudo-cycles visible around the stages of the early universe. While fundamentally local and topologically trivial, such deformations can provide non-trivial pseudo-homological effects clearly distinguishable over a large (albeit local) temporal region of the early universe. A distinction must be made manifest. While real axions are defined based on true cycles arising on the compactified directions, the pseudoaxions arise on deformations of the underlying manifold that do not require compactification to begin with. 
 In terms of enumerative geometry, the number of cycles allowed by a Calabi-Yau manifold is usually large. The simplest Calabi Yau manifold, the six-torus, will provide us with $(6\times 5)/2=15$ different two-cycles and the same amount of four-cycles. If we want to go to a more complicated compactification capable of providing us the Standard Model at low enough energies, the number of possible two-cycles may rise up to one hundred. It is however important to note that this number, while dependent on the chosen geometry, and usually relatively large, is finite. In the case of possible pseudo-cycles, the number is limited by the pseudo-homology and the allowed relative topologies. Again, this number will also be finite.
 The main object to be analysed is a throat with a capped end which generates an extension towards lower energies but which ends at energies corresponding to the QCD scale in the early universe. The additional parameter being induced is the length of the throat, which governs the scale at which the effect of the pseudo-axions ends. This is associated to a relative increase in apparent "dark energy" effects which compensate for the observed $H_{0}$ tension. This article deals with string phenomenology as it tries to connect the perturbative ten-dimensional effective field theories describing the massless degrees of freedom of string theory at very high energy (for example type IIA-IIB string theory with D-branes) and the low energy phenomena of the emerging four-dimensional universe. Indeed, the phenomena discussed are in a sense at an intermediate range between these regions, the $H_{0}$ tension observed nowadays in cosmology being expected to be a result of such intermediate scale phenomena. Moduli stabilisation is a particularly important aspect of string phenomenology, implying that expectation values of moduli fields determine many parameters of the low energy effective field theory. Moreover, such expectation values also parametrise the shape and size of the extra dimensions leading to a fascinating connection between the string world and the four dimensional cosmology. Gauge couplings or Yukawa couplings arising in the low energy domain are determined by such moduli expectation values. Quantum corrections often arise in order to fix such expectation values resulting in non-zero masses for their particle excitations. The standard model of elementary particles is expected to exist as a realisation of a stack of spacetime filling branes wrapping cycles in the compact dimensions. Gravity on the other side is expected to propagate in the bulk leading to a string scale $M_{s}\sim M_{Planck}/\sqrt{V}$, of course, with a large compactification volume $V$. While real axion fields appear as Kaluza-Klein zero modes of the ten-dimensional form fields, the pseudo-axions I will consider here are only visible in the intermediate region, as they arise as modes over pseudo-cycles which are fundamentally depending just on the geometry and pseudo-homology as visible in the intermediate and high energy (stringy) region. 
 Focusing, for the sake of exemplification, on the type $IIB$ string theory we consider the Calabi-Yau manifolds for the additional spatial dimensions. Fluxes in conformally flat six-dimensional spaces can break $N=4$ supersymmetry down to $N=3,2,1,0$ in a well defined and stable way. While doing so, fluxes also give vacuum expectation values to several moduli fields arising from the compactification. However, fluxes give rise to positive contributions to the energy momentum tensor and in order to compensate for that we need some sources for negative tension. These sources are usually orientifold planes. The standard fields are the dilaton $\phi$, the metric tensor $g_{MN}$ and the antisymmetric $2$-tensor $B_{MN}$ in the NS-NS sector. The massless RR sector contains $C_{0}$, the $2$-form potential $C_{MN}$ and the four-form field $C_{MNPQ}$ with the self-dual five-form field strength. We can combine the two scalars $C_{0}$ and $phi$ into a complex field $\tau=C_{0}+ie^{-\phi}$ parametrising the $SL(2,\mathbb{R})/U(1)$ space. The fermionic superpartners are two Majorana-Weyl gravitinos  of the same chirality $\gamma_{11}\psi^{A}_{M}=\psi^{A}_{M}$ and two Majorana-Weyl dilatons $\lambda^{A}$ with opposite chirality with respect to gravitinos. The field strength for the NS flux is $H=dB$ and for the RR field strengths, we have 
 \begin{equation}
 F^{(10)}=dC-H\wedge C + me^{B}=\hat{F}-H\wedge C
 \end{equation}
 where of course $\hat{F}=dC+me^{B}$. The RR fluxes are constrained by the Hodge duality
 \begin{equation}
 F_{n}^{(10)}=(-1)^{[n/2]} \star F_{10-n}^{(10)}
 \end{equation}
 The Bianchi identities are
\begin{equation}
\begin{array}{cc}
dH=0, &  dF^{(10)}-H\wedge F^{(10)}=0
\end{array}
\end{equation}
Sources will clearly alter the potentials leading to no globally well defined potential. Integrating the field strength over a cycle in the presence of sources does not necessarily result in a null outcome. This situation implies the existence of a non-zero flux. As charges are generally quantised in string theory, the fluxes will also be constrained by Dirac quantisation prescriptions. Given the Bianchi identities, the fluxes will satisfy 
\begin{equation}
\frac{1}{(2 \pi\sqrt{\alpha '})^{p-1}}\int_{\Sigma_{p}}\hat{F}_{p}\in \mathbb{Z}
\end{equation}
for a $p$-cycle $\Sigma_{p}$. The number of $2$ and $4$ cycles are identical due to hodge duality. The 3-cycles appear as pairs. 
The axion field arises because of the reduction of a NS-NS two form $B_{2}$ on a two-cycle with continuous shift symmetry. This appears as a result of some higher dimensional gauge symmetry of the two-form. Including branes and fluxes we also obtain a monodromy which has been broadly discussed in [15] and [16].

 As an example we can consider a $D5$ brane wrapping an internal two-cycle. We obtain the axion as $b=\frac{1}{\alpha '}\int_{\Sigma_{2}}B_{2}$ with the classical shift symmetry $b\rightarrow b + const$. The $D5$ brane breaks the shift symmetry and moreover allows a monodromy which can give rise to monodromy driven inflation. Taking a look at the Dirac-Born-Infled action for the $D5$ brane
\begin{equation}
S_{D5}=\frac{1}{(2\pi)^{5}g_{s}\alpha'^{3}}\int_{\mathcal{M}\times \Sigma_{2}}d^{6}x\sqrt{-det(G_{ab}+B_{ab})}
\end{equation}
which is the effective field theory associated to string theory at lower energies. Integrating over the $\Sigma_{2}$ cycle one obtains the potential of the axion in the four dimensional effective theory
\begin{equation}
V(b)=\frac{\rho}{(2\pi)^{6}g_{s}\alpha'^{2}}\sqrt{(2\pi)^{2}l^{4}+b^{2}}
\end{equation}
where $l$ represents the size of the two-cycle $\Sigma_{2}$ in string units, while $\rho$ is the dimensionless coefficient generated by the warp factor. Various types of inflation have been analysed using the monodromy generated by D-branes. For example in ref. [16] the potential for the inflaton field has been constructed as a power law with the exponent depending on the integral over the internal manifold and the contributions from a Chern-Simons term. As has been observed in [16] additional flexibility can be gained by uncertainties in the integration over the internal space. 
Let us however now consider integration over pseudo-cycles arising in the internal space. These will also give rise to axion fields which however will disappear once the scale of validity of the identification between real homology and pseudo-homology is reached. There is no much difference between the two approaches until the critical point is reached, except for the size of the two cycle (considering the above example) which would be decreasing as a function of time. One must also take into account that the proliferation of initial pseudocycles will depend on the fluctuations in the initial manifold. 

It can be verified as a theorem (for proof see [20]) that every pseudocycle $f:V\rightarrow M$ of dimension $k$ induces a well defined integral homology class $\alpha_{f}\in H_{k}(M; \mathbb{Z})$. Also, any singular cycle $\alpha\in Z_{k}^{sing}(M;\mathbb{Z})$ gives rise to a $k$-pseudocycle $f: V\rightarrow M$ such that $\alpha_{f}=\alpha$. Therefore integral cycles in singular homology can be represented by pseudo-cycles. This implies that in the region of the early universe, we have a statistical ensemble of indistinguishable cycles and pseudocycles detectable either by singular homology defined over the early domain or by pseudohomology capable of detecting such pseudocycles and associating them to corresponding invariants. Consider a topological pair $(M,D)$ and a domain $S=CP^{1} - \{0\}$ defined by a punctured sphere with a marked point $z_{0}=\infty$. Take also a cylindrical end near $z=0$ given by $(s,t)\rightarrow e^{s+it}$. Out coordinate system $(s,t)$ can extend to all of $S - z_{0}$. We can consider a one-form $\beta$ which restricts to $dt$ on the cylindrical end and to zero on a neighbourhood of $z_{0}$. We may consider a non-negative, monotone non-increasing cutoff function $\rho(s)$ which is zero for $s$ much larger than zero and one for $s$ much smaller than zero. In this case we can write $\beta=\rho(s)dt$. Considering a fixed pair of complex structures $J_{0}\in J_{c}(M,D)$ together with $J_{F}$ we denote the space of complex structures by $J_{S}(J_{0},J_{F})$ giving $J_{S}\in C^{\infty}(S, J_{c}(M,D))$ in such a way that in a neighbourhood of $z_{0}$ we have $J_{S}=J_{0}$ and along the negative strip end we have $J_{S}=J_{F}$. In the neighbourhood of $z_{0}$ we also fix a distinguished tangent vector pointing in the positive real direction. For any element $\alpha\in H_{*}(\bar{X},\partial\bar{X})$, we can fix a relative pseudocycle representative $\alpha_{c}$ in such a way that $\partial\alpha_{c}\subset \partial \bar{X}$. Given such a pseudocycle and given an orbit $x_{0}$ in $\chi(M, H_{\lambda, t})$ we can choose a surface dependent almost complex structure $J_{S}\in J_{S}(J_{0}, J_{F})$. Given any orbit $x_{0}$ of this type we may define $\mathcal{M}_{M}(x_{0})$ as the space of the possible solutions to the map $u:S\rightarrow M$ satisfying 
\begin{equation}
(du - X_{H_{\lambda,t}}\otimes \beta)^{0,1}=0
\end{equation}
satisfying the asymptotic condition 
\begin{equation}
\lim_{s\rightarrow -\infty} u(\epsilon(s,t))=x_{0}
\end{equation}
Given any $x_{0} \in \chi(X; H_{\lambda,t})$ we can consider those $u\in \mathcal{M}_{M}(x_{0})$ in such a way that $u(S)\subset X$. This moduli space can be called $\hat{\mathcal{M}}_{M}(x_{0})$ and 
\begin{equation}
\mathcal{M}(\alpha_{c},x_{0})= \hat{\mathcal{M}}_{M}(x_{0})\times_{ev_{z_{0}}}\alpha_{c}
\end{equation}
Therefore we can define an extended moduli space described not only by the parameters that remain valid at later times, but also by those included in the pseudohomological description. This results in an extended space which includes several cycles over which we can integrate in a stable way (at least stable at early times). Integration towards the tip of this geometric construction will encounter conic singularities which can be avoided by allowing a smooth cut-off. This will allow a smooth, well behaved overall spacetime at later times. This technical aspect has been discussed in various references on orbifold compactification. I will not insist on these aspects here but of course the reader can look up references [17,18].

What interests us from a phenomenological point of view is an additional number of axions arising at high energies which will have an effect on the cosmological parameters. 
Returning to the four dimensional effective theory, we have the general effective action as
\begin{equation}
S_{4}= -\frac{1}{2}\int d^{4}x \sqrt{g_{4}}(\partial_{\mu}\phi \partial^{\mu}\phi + m^{2}\phi^{2})
\end{equation}
where the inflaton field $\phi$ becomes a variable in the inflaton mass, i.e. the inflaton mass varies along the inflaton field. This all implies the existence of a critical value $\phi_{c}$ which plays the role of a scale which triggers a jump in the potential term of the form 
\begin{equation}
m^{2}\phi^{2}=[m_{a}^{2}+(m_{b}^{2}-m_{a}^{2})\theta (\phi-\phi_{c})]\phi^{2}
\end{equation}
with $m_{a}$ and $m_{b}$ being the inflaton masses when the inflaton field is larger and respectively smaller than the critical value $\phi_{c}$. 

\begin{figure}
  \includegraphics[width=\linewidth]{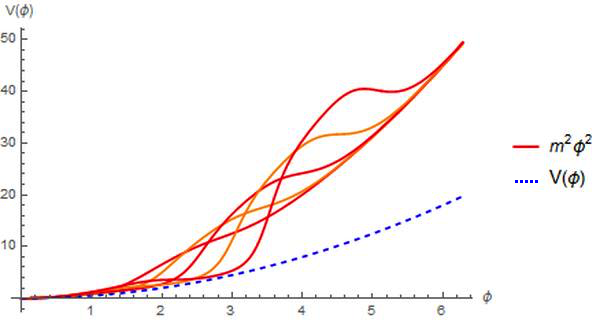}
  \caption{The potential computed with pseudocycle corrections (continuous curves) and without (dotted curve)}
  \label{graph}
\end{figure}

The discussion about the effects of the axion monodromy and its extensions due to D-branes wrapping torsion cycles have been discussed in ref. [15], [16]. I suggest here a similar approach with the distinction that there will be a certain number of pseudo-cycles generating non-trivial pseudo-homology leading to additional terms. 
Following the model for axion monodromy driven inflation by means of torsional cycles presented in [16] we expand the analysis towards pseudo-cycles to be found in the early stages. With the usual $N$ D3-branes along the $3+1$ dimensional space-time and the natural warp factor $e^{-2A}\sim 1/N$ we have the ten dimensional metric
\begin{equation}
ds_{10}^{2}=e^{-2A(y^{i})}dx_{\mu}dx^{\mu}+e^{2A(y^{i})}dy_{i}dy^{i}
\end{equation}
where the greek indices count the spacetime coordinates while the latin ones count the coordinates of the internal manifold. Disregarding certain aspects related to the warping, in a Calabi-Yau compactification the massless modes are in one-to-one correspondence with the elements of the cohomology group of the internal manifold. This observation is again modified by the presence of D-branes wrapping torsional cycles as presented in [16]. Working with the same relevant $\Sigma_{1}$ and $\Sigma_{2}$ cycles as in [16] here we will have to consider the fact that pseudo-cycles will contribute as well. The massless modes are not capable of detecting D-branes wrapping torsional cycles. Designating the unwrapped internal manifold with $X_{6}$ and considering the torsion cycles designated above, we consider the Laplacian eigenforms of $X_{6}$ in a similar way, namely 
\begin{equation}
\begin{array}{cc}
d\gamma_{1}=p\rho_{2}, & d\tilde{\rho}_{4}=p\tilde{\gamma}_{5}\\
d\eta_{2}=k\omega_{3}, & d\tilde{\omega_{3}}=k\tilde{\eta}_{4}\\
\end{array}
\end{equation}
where $\gamma_{1}$, $\tilde{\rho}_{4}$, $\eta_{2}$ and $\tilde{\omega}_{3}$ are the generators of $Tor(H_{i}(X_{6},\mathbb{Z}))$ for $i=1,4,2,3$. $\rho_{2}$, $\tilde{\gamma}_{5}$, $\omega_{3}$ and $\tilde{\eta}_{4}$ are trivial in the de Rham cohomology but non-trivial generators of the $H^{i}(X_{6},\mathbb{Z})$ group for $i=2,5,3,4$. We can expand in terms of this eigenforms and obtain 
\begin{equation}
\begin{array}{c}
B_{2}=b\eta_{2}+\bar{b}\rho_{2}+b_{1}\wedge \gamma_{1}+b_{2}\\
C_{2}=c\eta_{2}+\bar{c}\rho_{2}+c_{1}\wedge \gamma_{1}+c_{2}\\
C_{4}=\mathfrak{c}\tilde{\rho}_{4}+\bar{\mathfrak{c}}\tilde{\eta}_{4}+\mathfrak{c}_{1}\wedge\tilde{\omega}_{3}+\bar{\mathfrak{c}}_{1}\wedge\omega_{3}+\\
+\mathfrak{c}_{2}\wedge\rho_{2}+\bar{\mathfrak{c}}_{2}\wedge \eta_{2}+\mathfrak{c}_{3}\wedge\gamma_{1}+\mathfrak{c}_{4}
\\
\end{array}
\end{equation}
The distinction with respect to [16] is that here we must also include the pseudo-cycles giving rise to a different form of the potential term. 
Indeed, we obtain the four dimensional action 
\begin{equation}
S_{4}=-\frac{1}{2}\sqrt{g_{4}}(\partial_{\mu}\phi\partial^{\mu}\phi+m^{2}\phi^{2})
\end{equation}
and we also will have a critical inflaton field occurring around inflation, $\phi_{c}$. The mass term 
\begin{equation}
m^{2}\phi^{2}=[m_{a}^{2}+(m_{b}^{2}-m_{a}^{2})\theta(\phi-\phi_{c})]\phi^{2}
\end{equation}
with $m_{a}^{2}$ and $m_{b}^{2}$ the inflaton masses when inflaton field is smaller respectively larger than the critical value, will have a different structure, involving the statistics of a finite (albeit potentially large) number of pseudo-cycles modifying the expected potential term
\begin{equation}
V(\phi)=\frac{1}{2}m_{a}^{2}\phi^{2}+\frac{(m_{b}^{2}-m_{a}^{2})\phi^{2}}{2(1+exp(C_{H}(\phi^{2}-\phi_{c}^{2})/M_{Planck}^{2}))}
\end{equation}
Here the parameter $C_{H}$ is particularly important in describing the phenomenon of pseudo-cycle annihilation. This process can be described in the following way. After the cylindrical end extends enough towards lower energies, it will start folding upon itself leading to a decay geometry leading eventually to some form of cap at low enough energies. As stated before the associated moduli space will decay whenever the approximation leading to $\hat{\mathcal{M}}_{M}(x_{0})\times_{ev_{z_{0}}}\alpha_{c}$ cannot be sustained anymore. For the sake of simplicity I will consider this process as being governed by a gaussian distribution, modulated by the fact that there is a maximal (albeit large) number of possible pseudo-cycles supported by the geometry. As the pseudo-cycles disappear, the resulting geometry shifts towards one that favours accelerated expansion, after a phase of re-heating which can be seen in Figure \ref{graph}. 

\begin{figure}
  \includegraphics[width=\linewidth]{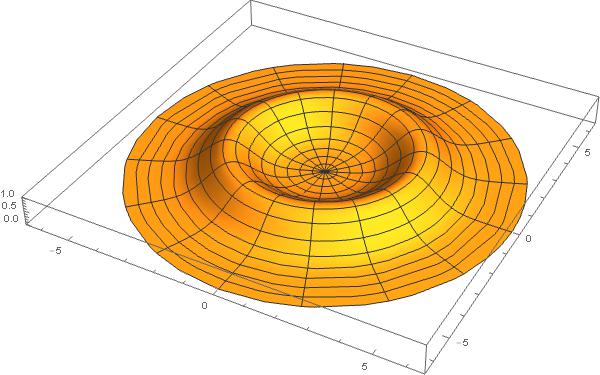}
  \caption{Geometric representation of the manifold on which pseudo-cycles are being considered, visible at sufficiently high energy and identifiable with real cycles. They can be detected by corresponding pseudo-homologies defined on the region of high energy. They decay as time increases and the energy decreases, i.e. when the distortion of the underlying manifold becomes negligible}
  \label{graph1}
\end{figure}
The pseudo-homology identifies a non-trivial component on which non-trivial pseudo-cycles can develop. Its defining parameter is given by the amplitude of the deformation above the flat background, a measure that is decaying as time advances. The construction has been represented in Figure \ref{graph1}. As the parameter $C_{H}$ measures the smoothness of the (pseudo)axion field and the associated potential, it is strongly dependent on the counting of pseudo-cycles in the early universe. This counting can be done by means of invariants capable of detecting them, and represents a calculation performed for example in [19]. Given a moduli space associated to the deformation we introduced, the parameter presented previously in the construction of the extension of the moduli space must be included in the definition of $C_{H}$. Indeed, once the approximation only sees the cylindrical end, the form $\beta$ restricts to $2 dt$ and the integration follows the method of the standard approach. However, the number of seen axion cycles is substantially increased, being considered as finite, while large, and the resulting cycles as indiscernible given the domain of the approximation. With this in mind we derive a formula for $C_{H}$ playing also the phenomenological role of a continuation function linking the early and the late universe in the form
\begin{equation}
C_{H}(\phi,\phi_{c})=\frac{exp(-(\phi -\phi_{c})^{2})} {(1+exp(\frac{\phi-\phi_{c}}{\phi_{c}}))}
\end{equation}
In this way we take into account the proliferation of pseudo-cycles in the initial phase, expanding the moduli space accordingly and allowing for new axion like particles in the early universe, while taking into account their dissipation at a later phase. The maximum number of pseudo-moduli accepted, while large, is finite. Its finite nature will play an important role in the end-stage of the pseudo-cycle proliferation, leading to a re-heating phase controlled by $\phi_{c}$ and by the length of the cylinder section of the distortion. 

\section{conclusion}
The problem of a discrepancy between early and late Hubble constant behaviour has been analysed from the perspective of the axion field and its potential. An additional element has been integrated, namely the fact that in the early stages of cosmological evolution, pseudo-cycles in the underlying manifold could contribute to additional axion cycles visible only at the very early times. These would decay once the pseudo-homology became trivial homology and the pseudo-cycles would vanish in the background geometry. However, they would account for additional particle production and additional contributions towards the re-heating phase. Therefore, not only would such pseudo-cycles offer new computational flexibility required for the explanation of the early vs. late Hubble constant discrepancy, but they would also add new contributions to the re-heating phase of the universe. 

\end{document}